\documentclass[twocolumn,twocolumn]{IEEEtran}
\usepackage[T1]{fontenc}
\usepackage{color}
\usepackage{float}
\usepackage{mathrsfs}
\usepackage{amsmath}
\usepackage{amssymb}
\usepackage{graphicx}
\usepackage{verbatim}
\usepackage[unicode=true,
 bookmarks=true,bookmarksnumbered=true,bookmarksopen=true,bookmarksopenlevel=1,
 breaklinks=false,pdfborder={0 0 0},pdfborderstyle={},backref=false,colorlinks=false]
 {hyperref}
\hypersetup{pdftitle={Your Title},
 pdfauthor={Your Name},
 pdfpagelayout=OneColumn, pdfnewwindow=true, pdfstartview=XYZ, plainpages=false}

\makeatletter

\floatstyle{ruled}
\newfloat{algorithm}{tbp}{loa}
\providecommand{\algorithmname}{Algorithm}
\floatname{algorithm}{\protect\algorithmname}

\usepackage[caption=false,font=footnotesize]{subfig}
\usepackage{algorithm}
\usepackage{algorithmic}

\usepackage{multirow} 
\usepackage{amsmath} 
\usepackage{xcolor}

\allowdisplaybreaks[4]

\ifCLASSOPTIONcompsoc
\usepackage[caption=false,font=normalsize,labelfont=sf,textfont=sf]{subfig}
\else
\usepackage[caption=false,font=footnotesize]{subfig}
\fi

\usepackage{cite}
\usepackage{bm}
\usepackage{algorithmic}
\usepackage{algorithm}
\usepackage{graphicx}
\IEEEoverridecommandlockouts

\usepackage{lettrine}

\usepackage{geometry}
\@ifundefined{showcaptionsetup}{}{%
 \PassOptionsToPackage{caption=false}{subfig}}
\usepackage{subfig}
\makeatother

\begin{document}

\title{\textcolor{black}{Joint UAV Trajectory Planning and
LEO Satellite Selection for Data Offloading in Space-Air-Ground Integrated Networks
}}

\author{\IEEEauthorblockN{Boran Wang$^{1}$, Ziye Jia$^{1}$, Can Cui$^{1}$, and Qihui Wu$^{1}$ \\}
\IEEEauthorblockA{$^{1}$The Key Laboratory of Dynamic Cognitive System of Electromagnetic Spectrum Space, Ministry of Industry and Information Technology, Nanjing University of Aeronautics and Astronautics, Nanjing, Jiangsu, 211106, China\\
\{wangboran, jiaziye, cuican020619, wuqihui\}@nuaa.edu.cn}

\thanks{{This work was supported in part by National Natural Science Foundation of 
China under Grant 62301251, in part by the Natural Science Foundation of Jiangsu 
Province of China under Project BK20220883, in part by the Aeronautical Science 
Foundation of China 2023Z071052007, and  in part by the Young Elite Scientists 
Sponsorship Program by CAST 2023QNRC001
}( \textit{Corresponding author: Ziye Jia}).} }

\maketitle
\pagestyle{empty} 
\maketitle
\pagestyle{empty} 

\thispagestyle{empty}
\begin{abstract}
    With the development of low earth orbit (LEO) satellites and unmanned aerial vehicles (UAVs), the space-air-ground integrated network (SAGIN)
    becomes a major trend in the next-generation networks. 
    However, due to the instability of heterogeneous communication and time-varying characteristics of SAGIN, 
    it is challenging to meet the remote Internet of Things (IoT) demands for data collection and offloading. In this paper, 
    we investigate a two-phase hierarchical data uplink 
     model in SAGIN. Specifically, UAVs optimize trajectories to enable efficient data collection
     from IoT devices, and then they transmit the data 
    to LEO satellites with computing capabilities for further processing.
    The problem is formulated to minimize the total energy consumption for IoT devices, UAVs, and LEO satellites. 
    Since the problem is in the form of mixed-integer nonlinear programming and intractable to solve directly,
    we decompose it into two phases. In the IoT-UAV phase, we design the algorithm to jointly 
    optimize the IoT pairing, power allocation, and UAVs trajectories. 
    Considering the high dynamic characteristics of LEO satellites,  
    a real-time LEO satellite selection mechanism joint with the Satellite Tool Kit is proposed in the UAV-LEO phase.
    Finally, simulation results show the effectiveness of the proposed algorithms, with about 10$\%$ less
    energy consumption compared with the benchmark algorithm.
\end{abstract}
\newcommand{\CLASSINPUTtoptextmargin}{0.8in}

\newcommand{\CLASSINPUTbottomtextmargin}{1in}

\section{Introduction}
\lettrine[lines=2]{T}{he} Internet  of Things (IoT) devices are widely applied in the daily life, such as environmental monitoring and 
traffic management. 
 However, due to the limited ground base stations in remote or post-disaster areas, it is difficult to satisfy the 
demands for data collection and offloading supported by the terrestrial networks. 
The space-air-ground integrated network (SAGIN)
is perceived as an effective solution to tackle the above
difficulties \cite{9184929}. 
In SAGIN, low earth orbit (LEO) satellites can provide the IoT devices with extensive connectivities \cite{10745905,9933792}. 
Additionally, the in-orbit computing allows LEO satellites to directly process tasks, which avoids the long 
propagation delays and eases the congestion on 
bandwidth-limited downlink channels \cite{11006480,9760009}. 
Moreover, unmanned aerial vehicles (UAVs), as ideal candidates for aerial relays, 
can be deployed flexibly to ensure efficient data collection \cite{10778646}. 
On one hand, the UAVs trajectories can be optimized to minimize the 
multi-hop transmission and propagation distance\cite{9307224}. Besides, UAVs facilitate the 
line-of-sight (LoS) communications with ground devices for a wide view, improving the 
channel quality and enhancing the transmission throughput \cite{9454514,8660516}. Nevertheless, the limitation of communication resources
restricts the number of IoT devices served by UAVs and
leads to a poor spectrum efficiency\cite{7315054}.
In response to this issue, the non-orthogonal multiple 
access (NOMA) technology, which emerges as a promising paradigm, 
allows multiple IoT devices to share a single resource block. 
\par  Some works have begun to explore problems on resource allocations in SAGIN.
 The authors in \cite{9839554} propose an iterative power allocation algorithm
to maximize the sum rate in a NOMA-based hybrid satellite-UAV-terrestrial network. In \cite{10899883}, the authors 
consider the complexity of SAGIN and sovle the service function chain scheduling problem by incorporating
deep reinforcement learning.
The authors in \cite{10440193} study the total energy consumption minimization for task
processing in an SAGIN-supported mobile edge computing system. In \cite{10454605}, 
the authors introduce a data collection scheme to balance the throughput and fairness 
among the IoT nodes in SAGIN. Although the above works are conducted in SAGIN, 
the satellite selection issues are not considered, which can significantly 
enhance the performance of the system.
\begin{figure}[htbp]
    \centering
    \includegraphics[width=7.5cm]{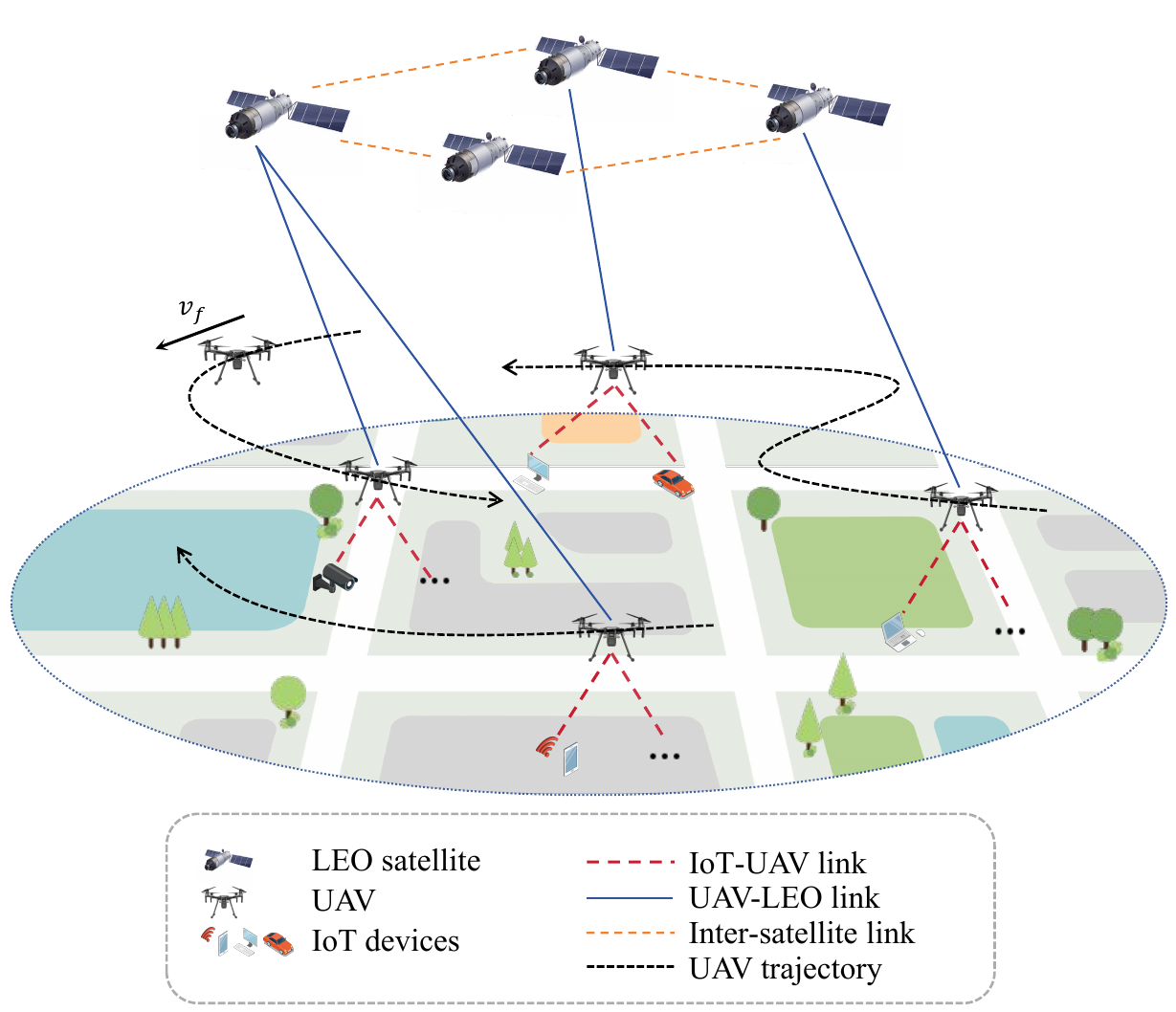}
    \caption{SAGIN model including UAV-based data collection and LEO satellite offloading selection. }
    \label{fig:aa}
    \end{figure}
\par Considering the instability of heterogeneous communications and
 time-varying characteristics of SAGIN, we propose 
a hierarchical framework incorporating IoT devices on the ground, UAVs serving as aerial relays, and LEO 
satellites with computing capabilities. The problem is formulated to minimize the total energy consumption. 
To tackle the problem, the total process is divided into two phases. In the first phase, we design 
the algorithm for the  IoT pairing, power allocation, and UAV trajectory planning. 
In the second phase, we develop a LEO satellite selection mechanism joint with Satellite Tool Kit (STK). 

\par The remainder of this paper is organized as follows. In
Section \ref{sec:System-Model}, we design the system model and provide the
problem formulation. In Section \ref{sec:Algorithm-Design}, the algorithms are proposed. Section \ref{sec:simulation} evaluates the performance of the proposed
algorithms via numerical analyses. Finally, the conclusions are drawn in Section \ref{sec:conclusion}.


\section{System Model\label{sec:System-Model}}

As shown in Fig. \ref{fig:aa}, we consider an SAGIN 
which consists of $U$ UAVs denoted by 
$\mathcal{U}=\left[ 1,2,\cdots ,U\right]$, and $S$ LEO satellites 
indicated by $\mathcal{S}=\left[ 1,2,\cdots ,S \right]$. 
In addition, $K$ IoT devices scattered randomly on the
ground are represented as $\mathcal{K}=\left[ 1,2,\cdots ,K \right] $.
Since IoT devices have limited computing capabilities, they need to transmit their data to LEO satellites 
for processing. However, the limited transmission power of IoT 
devices results in the inability to directly transmit data towards
LEO satellites. Therefore, UAVs, served as aerial base stations, 
are deployed in the area for data collection from IoT devices and forward the data to LEO satellites for further computing.
Consequently, we divide the process of data uploading into two phases. 
In the first phase, the UAVs collect data from the IoT devices orderly. 
In the second phase, UAVs select LEO satellites for computational offloading
after they finish the data collection at a certain hover position.

\subsection{Data Collection from IoT to UAV}
\par In the IoT-UAV phase, the NOMA technology is adopted to
improve the transmission efficiency. We assume that two device nodes within a certain 
range can form a pair, and these paired nodes follow the
NOMA technology when transmitting data towards UAVs. 
For the unpaired nodes, the orthogonal frequency division multiple 
access (OFDMA) technology is employed for data transmission.
Besides, the binary variable
$\alpha _{k,m}\in \left\{ 0,1 \right\}$ is defined to represent the relationships among IoT devices. 
Specially, $\alpha _{k,k}=1$ indicates the $k$-th IoT device 
is an independent device node and transmits its data to the corresponding UAV via the OFDMA technology. If $\alpha _{k,m} = 1$ and
$k\neq m$, the $k$-th
IoT device is associated with $m$-th IoT device for data transmission 
by the NOMA technology during the whole mission time, 
and the signal from the $m$-th device node is decoded later than that from
the $k$-th device node. 
\par The three-dimensional Cartesian coordinate
 is considered to describe the locations of UAVs and IoT devices. 
The IoT devices are distributed randomly on the
ground, and the horizontal coordinates of the $k$-th IoT device 
are denoted by $\mathbf{q}_{k}=\left( x_k,y_k,0 \right) $. 
All UAVs have a fixed flying height $h_u$,  
and the horizontal plane coordinate of the $u$-th UAV at the {\it n}-th hover point 
is denoted as ${\mathbf{q}_u\left(n\right)=\left( x_{u}\left(n\right),y_{u}\left(n\right),h_u \right)} $.
The communication link from IoT devices to UAVs can be approximated as an LoS link, and the channel gain from 
the IoT device $k$ to the corresponding UAV is
\begin{equation}
    G_{k}=\frac{\beta_0}{h_{u}^{2}+\left( x_{u}\left( n \right) -x_k \right) ^2+\left( y_{u}\left( n \right) -y_k \right)^2}, 
\end{equation}
where ${\beta_0}$ denotes the channel gain at the reference distance $r_0=1\mathrm{m}$.
 Furthermore, when decoding the signal from IoT device $k$, 
the signal from IoT device $m$ is regarded as the noise. Hence, the received signal-to-interference-plus-noise 
ratio from the $k$-th IoT device is
\begin{equation}
    SINR_{k}=\frac{p_kG_k}{\sum_{m\neq k,m\in\mathcal{K}}\alpha_{k,m}p_mG_m+\sigma_{iu} ^2},
\end{equation}
\noindent where $p_k$ is the transmission power of the $k$-th IoT device, and ${\sigma_{iu} ^2}$ represents 
the noise power. Hence, the achievable data rate of the $k$-th IoT device is
\begin{equation}
  d_{k}=B_{iu}\log_{2}\left( 1+{SINR_{k}} \right),
\end{equation}
where ${B_{iu}}$ is the available bandwidth between the UAV and IoT device.   
The transmission delay for $k$-th IoT device is
\begin{equation}
    T_{k}^{tr}=\frac{{D_{k}}}{d_{k}},
\end{equation}
where ${D_{k}}$ denotes the data size
of task data packet that needs to be processed from IoT device $k$. 
The energy consumption of IoT device $k$  for transmitting data to
the associated UAV can be obtained by
\begin{equation}
    E_{k}=p_{k}T_{k}^{tr}.
\end{equation}
\par Besides, we assume that the UAV visits the NOMA groups and the independent
points in a certain order after UAV trajectory is planned. 
The trajectory of UAV $u$ can be expressed as $\left\{\mathbf{q_u}(0), \mathbf{q_u}(1), \cdots, \mathbf{q_u}(N_u)\right\}$, 
where $N_u$ represents the number of hover points of the $u$-th UAV. Hence, the trajectory length of UAV $u$ throughout the overall period can be defined as
 \begin{equation}L_u=\sum_{n=0}^{N_u}\left\|\mathbf{q}_u{\left(n+1\right)}-\mathbf{q}_u{\left(n\right)}\right\|.\end{equation}

\begin{figure}[htbp]
    \centering
  
    \subfloat[Positional relationship between the UAV and LEO satellite.]
    {\includegraphics[width=0.4\textwidth]{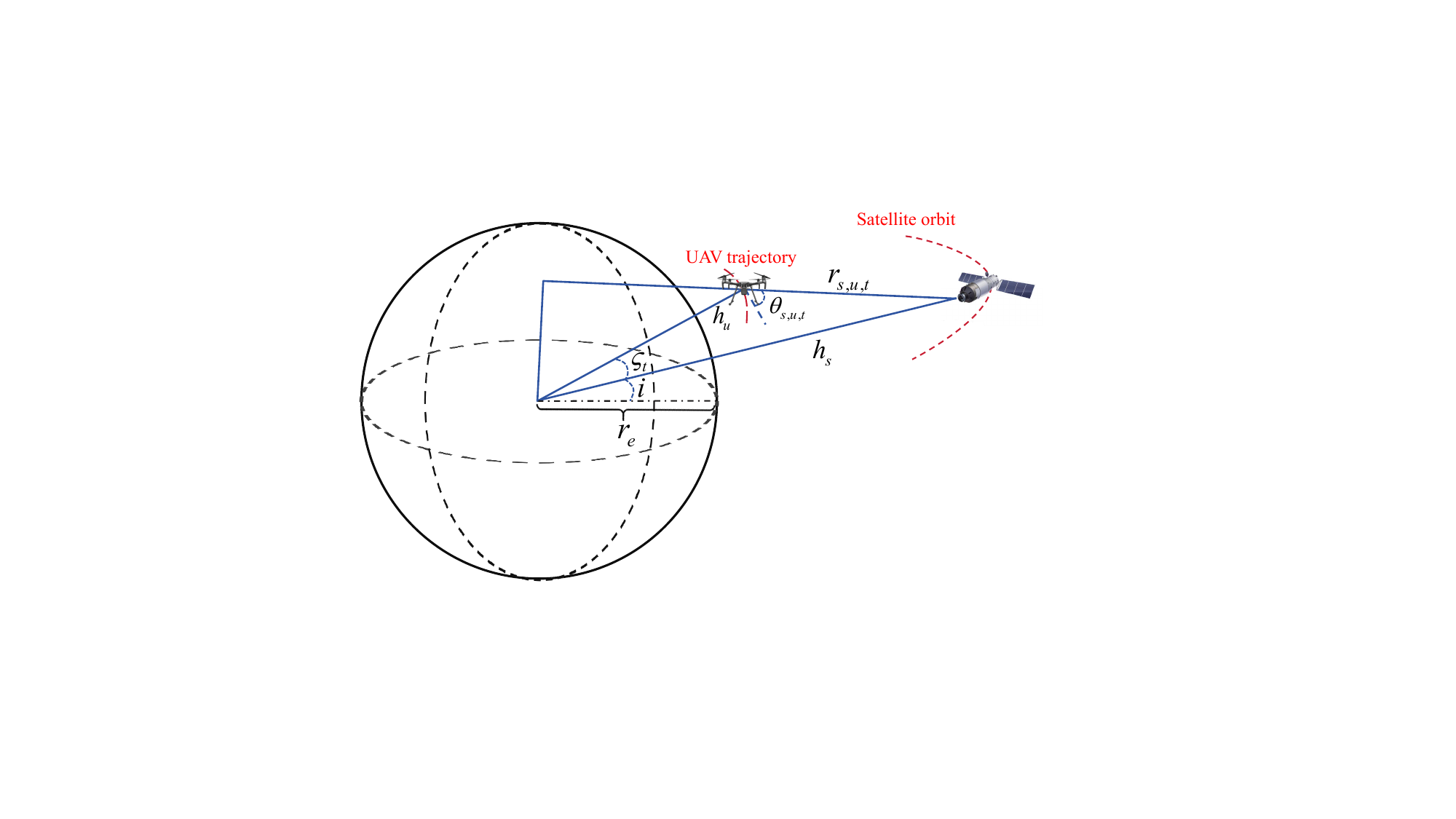}\label{fig:bb}}
    \hfill   
    \subfloat[Coverage area of the LEO satellite.]
    {\includegraphics[width=0.4\textwidth]{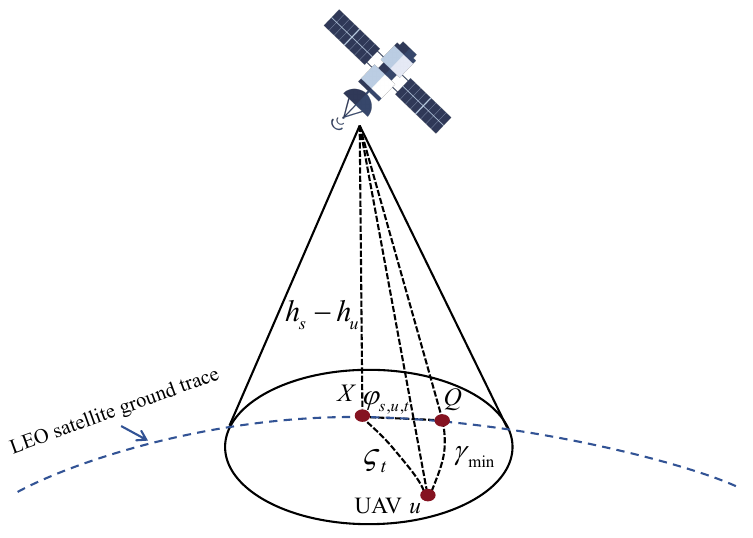}\label{fig:cc}}
    \hfill
    \caption{Geometrical representation of the UAV-LEO communication link.}
    \label{fig:mm}
  \end{figure}

The time duration for flying with the constant speed $v_f$ of the $u$-th UAV is
 \begin{equation}
    T_{u}^{fly}=\frac{L_u}{v_f}.
\end{equation}
Furthermore, the UAVs hover in the air when they collect data from IoT devices. 
Therefore, the hovering time duration for UAVs in the first phase mainly 
depends on the transmission delay of IoT devices, which can be given by
\begin{equation}
    \begin{split}
        T^{hov}_{iu}=&\sum_{k=1}^K\sum_{\substack{m=1\\m\neq k}}^K{\alpha _{k,m}}\left(T_{k}^{tr}+T_{m}^{tr}\right)+\sum_{l=1}^K{\alpha _{l,l}}T_{l}^{tr}.
    \end{split}
\end{equation}
\par Note that in the current phase, the energy consumption of UAVs is mainly composed of  
hovering and the execution of flight missions. The total energy consumption of both IoT devices and UAVs is
\begin{equation}E_{iu}^{total}=P_hT^{hov}_{iu}+\sum_{u=1}^{U}P_fT_{u}^{fly}+\sum_{k=1}^{K} E_{k},\label{eq:4} \end{equation}
where $P_h$ and $P_f$ are the hovering and flight power of the UAVs, respectively.

\subsection{Data Offloading from UAV to LEO  }
\par When the UAV receives all 
data from IoT devices, it obtains position information from 
all visible LEO satellites and selects the satellite with the maximum throughput.
Considering the high variability of LEO satellites, 
the association binary variable 
$\beta_{s,u,t}\in \left\{ 0,1 \right\}$ is introduced to represent the computation 
offloading decision of UAV $u$ at $t$.
 The channel gain between  
UAV $u$ and LEO satellite $s$ can be 
calculated as
\begin{equation}
    G_{s,u,t}=92.44\mathrm{dB}+\log_{10}(r_{s,u,t})+\log_{10}(f_s),
\end{equation}
where $f_{s}$ indicates the operating frequency, and $r_{s,u,t}$ represents the distance between the $s$-th LEO satellite 
and  $u$-th UAV at $t$. As shown in Fig. \ref{fig:mm}\subref{fig:bb}, $r_{s,u,t}$ can be expressed as
\begin{equation}
    \begin{split}
    r_{s,u,t}=&\sqrt{(r_e+h_{s})^2-(r_e+h_u)^2\cos(\theta_{s,u,t})^2}\\
    &-(r_e+h_u)\sin(\theta_{s,u,t}),
    \end{split}
\end{equation}
in which $r_e$, $h_u$ and $h_{s}$ denote the earth 
radius, UAV altitude and LEO satellite altitude, respectively. $\theta_{s,u,t}$ represents the $s$-th LEO satellite  
elevation angle towards $u$-th UAV at $t$. According to \cite{Seyedi2012},
$\theta_{s,u,t}$ is calculated as 
\begin{equation}\theta_{s,u,t}=\tan^{-1}\left(\frac{\cos\left(\varphi_{s,u,t}\right)\cos\left(\gamma_{\min}\right)-\left(\frac{r_{e}+h_{u}}{r_{e}+h_{s}}\right)}{\sin\left(\cos^{-1}\left(\cos\left({\varphi_{s,u,t}}\right)\cos\left(\gamma_{\min}\right)\right)\right)}\right),\label{elevation} \end{equation}

\noindent where $\gamma_{\min}$, as illustrated in Fig. \ref{fig:mm}\subref{fig:cc}, is the smallest angle for the 
angular distance between the ground trace of 
the LEO satellite and UAV. $\varphi _{s,u,t}$ is angle on the 
surface of the earth connecting sub-satellite points $X$ and  
$Q$, which is the closest point to UAV $u$, i.e.,  
\begin{equation}
    \varphi_{s,u,t}=\left(\omega_E\cos i-\omega_S\right)\left(t-t_0\right)+\varphi_{s,u,t_0},  
\end{equation}
where $\omega_E$, $\omega_S$ and $i$ are the 
angular velocity of the earth rotation, angular velocity of the LEO satellite and satellite inclination, respectively.
$t_0$ is the time when the LEO satellite becomes visible.
\par Furthermore, the received power of LEO satellite is 
defined as
\begin{equation}
    P_{re}=P_{tr}G_{re}G_{tr}\left(\frac{c}{4\pi f_{s}r_{s,u,t}}\right)^2,
\end{equation}
where $P_{tr}$ is the transmission power from the UAV to LEO satellite.  
$G_{tr}$ and $G_{re}$ denote the transmission antenna gain of the UAV 
and the receiver antenna gain of the LEO satellite, respectively.
$c$ is the speed of light. Hence, the data rate can be calculated as
\begin{equation}
    d_{s,u,t}=B_{su}\log_{2}(1+\frac{P_{re}G_{s,u,t}}{\sigma_{su}^2}),
\end{equation}
where $B_{su}$ is the available bandwidth between the UAV 
and LEO satellite. $\sigma_{su}^2$ is the noise power.
Thus, the transmission delay from UAV $u$ to LEO satellite $s$ is obtained by
\begin{equation}
    T_{s,u,t}^{tr}=\frac{D_{u}}{d_{s,u,t}}\label{eq:5} ,
\end{equation}
where $ {D_{u}}=\sum_{k=1}^K\sum_{m=1}^K\alpha _{k,m}\left({D_{k}+\mathrm{sgn}\left[|k-m|\right]D_{m}}\right) $ is the total 
received data of the UAV from IoT devices, and $\mathrm{sgn}\left[\cdot \right]$ is defined as
\begin{equation}\operatorname{sgn}\left[x\right]=
    \begin{cases}
    1, & \operatorname{if}x>0,\\
    0, & \operatorname{if}x=0, \\
    -1, & \operatorname{if}x<0. 
    \end{cases}\end{equation}
\par The computing delay of the IoT devices through LEO satellite 
offloading can be expressed as
\begin{equation}
    T_{s,t}^{sa}=\frac{D_{s,t}}{c^{sa}},
\end{equation}
in which $D_{s,t}=\sum_{u=1}^{U}\beta_{s,u,t}D_{u}$ is the total amount of data needed 
to be processed by the $s$-th LEO satellite, and $c^{sa}$ denotes the computing capability of the LEO satellites. 
Besides, the backhaul latency is ignored 
since the data size of computation results is small.
\par The UAVs hovering delay is
\begin{equation}
    T^{hov}_{su}=\sum_{s=1}^S\sum_{u=1}^U\sum_t{\beta _{s,u,t}}\left[T_{s,u,t}^{tr}+T_{s,t}^{sa}\right]. 
\end{equation}
\par As for LEO satellites, the entire energy budget is for processing the data packets. Thus, in this phase,
the total energy consumption of both UAVs and LEO satellites is 
\begin{equation}
    E_{su}^{total}=\sum_{s=1}^S\sum_t{P_sT_{s,t}^{sa}}+P_hT^{hov}_{su} \label{eq:6},
    \end{equation}
    where $P_s$ is a constant power for LEO satellite computing units to process data packets.
\par Consequently, the total energy consumption of IoT devices, UAVs and LEO satellites is
\begin{equation}
    E^{total}=E_{iu}^{total}+E_{su}^{total} \label{eq:7}.
\end{equation}

\subsection{Problem Formulation}

\par We aim to minimize the total energy consumption of the proposed hierarchical SAGIN model by jointly optimizing
the association variable $\boldsymbol{\alpha}=\left\{\alpha_{k,m},\forall k,\forall m\right\}$, $\boldsymbol{\beta}=\left\{\beta_{s,u,t},\forall s,\forall u,\forall t\right\}$,
power allocation $\boldsymbol{p}=\left\{p_{k},\forall k\right\}$ and the hovering position  as well as visiting order $\boldsymbol{q}=\left\{\mathbf{q}_{u}\left(n\right),\forall u\right\}$. 
The optimization problem is detailed as 

\begin{align}
P0:&\underset{\boldsymbol{q},\boldsymbol{\alpha},\boldsymbol{\beta},\boldsymbol{p}}{\min}\,\,{E^{total}} \nonumber \\
\textrm{s.t.}\,\,&p_{\min}\leqslant p_{k}\leqslant p_{\max}, \forall k,\label{cons2}\\
&\rho p_mG_m\leqslant \alpha_{k,m}p_kG_k, \forall k, m,  \label{cons3}\\
&\mathbf{q}_u{\left(N_u\right)}=\mathbf{q}_u{\left(0\right)},\forall u,\label{cons4}\\
&\theta _{s,u,t}>\theta _{\min}, \forall u, s,\label{cons7}\\
&\sum_{k=1}^K{D_{k}}=\sum_{s=1}^S\sum_t{D_{s,t}},\label{cons8}\\
&\sum_{s\in \mathcal{S}}^{}{\beta_{s,u,t}}\leqslant 1, \forall u, t,  \label{cons5}\\
&\beta_{s,u,t} \in \left\{ 0,1 \right\}, \forall u, s, t, \label{cons6}\\
&\alpha_{k,m} \in \left\{ 0,1 \right\}, \forall k, m. \label{cons1}
\end{align}

\noindent Constraint (\ref{cons2}) represents the limitation
on the transmit power of the IoT device. Constraint (\ref{cons3}) ensures that the signals
can be decoded at the receiver. $\rho$ is a pre-determined power difference ratio and $\rho<1$.
 Constraint (\ref{cons4}) guarantees UAVs return to their start locations after finishing the mission. 
 Constraint (\ref{cons7}) illustrates the elevation constraint 
for establishing connections between LEO satellites and UAVs. 
Constraint (\ref{cons8}) ensures all data are uploaded and processed.
Constraints (\ref{cons5}) and (\ref{cons6}) indicate each UAV 
transmits its collection data to at most one LEO satellite at any time.
Due to the coupled variables, $P0$ is a mixed-integer nonlinear programming problem, 
which is NP-hard and intractable to solve.
\section{Algorithm Design\label{sec:Algorithm-Design}}

\par  Since the proposed SAGIN model incorporates two phases, and 
the LEO satellite selection can be determined after the UAVs trajectories are planned, $P0$ can be
decoupled into two stages for effective solutions. In detail, 
Algorithm \ref{alg:UAV_Optimization} is performed to obtain the 
IoT pairing, power allocation, and 
 flight trajectories of UAVs in the data collection from IoT to UAVs.  
In the data offloading from the UAV to LEO, the LEO satellite
with the best link throughput is selected to
 reduce the energy consumption through Algorithm \ref{alg:cc}.

\subsection{UAV Data Collection and Energy Optimization}

\par In order to effectively solve problem $P0$, we firstly 
consider the total energy consumption $E_{iu}^{total}$ in the IoT-UAV phase.
 $\boldsymbol{\alpha}$ are determined based on the greedy method. 
Specially, the nearest two IoT device nodes are paired to quickly obtain the NOMA groups. 
 Then, the optimization 
problem of IoT power and UAV hovering points is a non-convex problem. 
As shown in Algorithm \ref{alg:UAV_Optimization}, we
decompose the problem, in which
the hovering positions of UAVs and the transmission power are solved alternately for each pair of nodes (lines 3-10).
\noindent In detail, with the fixed transmission power of one device, 
 the transmission power of the other device is updated via the Newton method in turn until reaching the convergence.
The hovering position of the UAV is updated by the Nelder-Mead method until the result converges. 
Moreover, UAVs directly hover above the unpaired IoT devices which adopt the OFDMA technology,
and the unpaired IoT devices transmit data with the maximum power $p_{\max}$ to maximize the throughput.
\par Based on the given hovering points, the UAV trajectory planning problem is
essentially transformed into a multiple traveling salesman problem. 
To solve the problem effectively, we employ the 
grey wolf optimization algorithm \cite{8660872}. In detail,
after generating initial solutions for UAV trajectories, the objective function values are calculated regarding the flight distance. 
Then, the top three best solutions are obtained and denoted as $\alpha$, $\beta$ and $\gamma$ wolves, respectively.
By updating the positions of the three wolves based on the fitness values until the convergence is achieved,
the best UAVs trajectories solution can be obtained.
\begin{algorithm}
    \caption{UAV Data Collection and Energy Optimization}
    \label{alg:UAV_Optimization}
    \begin{algorithmic}[1]
        \REQUIRE All IoT devices' positions $\mathbf{q}_{k}$ and corresponding packets $D_{k}$, 
            and UAV initial position $\mathbf{q}_u{\left(0\right)}$.
        \ENSURE Energy consumption $E^{total}_{iu}$.
        \STATE Initialize the distance matrix and iteration number $J$. 
        The two nearest device nodes form an NOMA pair under the distance constraint. 
        Subsequently, the lists of paired and unpaired nodes are obtained.    
        \FOR{each paired nodes $k$ and $m$ }
        \STATE Initialize the transmission power $p_{k}$, $p_{m}$, hover position 
        coordinate $\left(x_{u},y_{u},h_u\right)$ and set $j=0$.
        \WHILE { $j$ $\leqslant$ $J$ }
        \STATE Update $p_{k}^{\left(j\right)}$, $p_{m}^{\left(j\right)}$ by alternately 
        fixing variables to iterate using Newton method.
        \STATE Update $\left(x_{u}^{\left(j\right)},y_{u}^{\left(j\right)},h_u\right)$ 
        by Nelder-Mead method.
        \STATE $j$ = $j$+1.
        \ENDWHILE
        \ENDFOR
    \STATE Obtain UAVs trajectories based the gray wolf optimization algorithm and compute energy 
    consumption $E^{total}_{iu}$ based on Eq. (\ref{eq:4}).
        \FOR{each unpaired node and paired node}
            \STATE exchange with other paired nodes.
            \STATE Update UAVs trajectories and transmission powers of IoT devices. 
            \STATE Calculate  $\hat{E}^{total}_{iu}$ according to  Eq. (\ref{eq:4}).
            \IF{$\hat{E}^{total}_{iu}$ < $E^{total}_{iu}$}
                \STATE Update $E^{total}_{iu}$.
            \ENDIF
        \ENDFOR
    \end{algorithmic}
    \end{algorithm}

\par To obtain better NOMA groups,
an operation of exchanging device nodes and updating the matching pairs 
is introduced, as shown in Algorithm \ref{alg:cc} (lines 12-19). To be specific, 
 all the unpaired nodes are tried to traverse with each other to check 
if there are any paired nodes that can be recombined, and these nodes are repaired.
 The potential NOMA groups in the new set of unpaired nodes are preferred to be paired with high priorities.
 Then, the energy consumption is updated after the repairing operations, 
 and the IoT pairing relationships are selected with the lowest energy cost. 
Similarly, for two NOMA groups that are close to each other, 
the paired nodes can be exchanged to update the pairing relationship.
By performing the exchanging operations 
between NOMA pairs and isolated OFDMA nodes, more NOMA pairs and less  energy consumption are enabled for further optimization.

\subsection{LEO Satellite Selection Optimization}
 
\par After the UAVs finish collecting data from IoT devices, they will offload the data to LEO satellites.
In the UAV-LEO phase, we mainly focus on minimizing the energy consumption $E^{total}_{iu}$.  
According to Eqs. (\ref{eq:5}) and (\ref{eq:6}), since the better throughput enables the less transmission delay, the UAV selects 
the LEO satellite with the best link throughput to reduce energy consumption. The detailed mechanism for LEO satellite selection based on 
the throughput in real time is designed in Algorithm \ref{alg:cc}. 
Through the joint simulation with STK, the accessibility from UAVs to LEO satellites can be obtained, 
and the position information of LEO satellites are known to the UAVs.
Hence, the elevation angle $\theta_{s,u,t}$ between UAVs and LEO satellites can be calculated based on Eq. (\ref{elevation}).
If $\theta_{s,u,t}$ < $\theta_{min}$, the satellite is discarded due to the disgusting interference.
After obtaining all the acceptable LEO satellites which satisfy the requirements, 
the UAV selects the LEO satellite with maximum link throughput.

\begin{algorithm}[!h]
    \caption{LEO Satellite Selection Optimization}
    \label{alg:cc}
    \textbf{Input:} UAV positions and the time for accessing the LEO satellite.\\
    \textbf{Output:} Best LEO satellite.
    \begin{algorithmic}[1]

        \STATE Initialize STK scenario, set $\theta_{min}$ as the elevation angle threshold and 
        relevant parameters.
        \STATE Compute the  accessibility of LEO satellites to the UAVs in real-time.
        \STATE Retrieve the position information of all visible LEO satellites.
        \STATE Compute the elevation angle $\theta_{s,u,t}$ of the LEO satellite based on Eq. (\ref{elevation}).
        Determine the set of LEO satellites whthin the elevation angle threshold.
        \STATE Calculate the LEO satellite communication link throughput and 
         select the best one.
    \end{algorithmic}
\end{algorithm}
\begin{table}[!ht]
    \centering
    \caption{SIMULATION PARAMETERS}
    \label{table1}
    \begin{tabular}{|c|c|c|c|}
    \hline
    \textbf{Parameter} & \textbf{Value} & \textbf{Parameter} & \textbf{Value} \\ \hline
        $p_{\max}$ & 5W & $p_{\min}$ & 0.1W \\ \hline
        $P_h$ & 80W & $P_f$ & 240W \\ \hline
        $P_s$ & 200W & $P_{tr}$ & 10W \\ \hline
        $B_{iu}$ & 1MHz & $B_{su}$ & 10MHz \\ \hline
        $f_s$ & 20GHz & ${\beta_0}$ & $9.89 \times 10^{-5}$ \\ \hline
        $G_{tr}$ & 10dBi & $G_{re}$ & 30dBi \\ \hline
        $r_e$ & 6378km & $h_u$ & 200m \\ \hline
        $\theta _{\min}$ & $15^{\circ}$ & $\rho$  & 0.8 \\ \hline
        $\omega_E$  & 7.29rad/s &  $c^{sa}$ & 25Mbit/s \\ \hline
        $\sigma_{iu}^2$ & $10^{-18}$W & $\sigma_{su}^2$ & $4 \times 10^{-14}$W \\ \hline
    \end{tabular}
\end{table}

\section{Simulation Results\label{sec:simulation}}
\par In this section, we conduct simulations to evaluate 
the performance of the proposed algorithms.  
IoT devices are randomly distributed within a square area with a 
side length of 500 meters.  
Three UAVs, located in 15$^{\circ}$N, 118$^{\circ}$E, 
start moving from the center of the area at a 
certain moment. 
200 LEO satellites are randomly selected from Starlink. 
The main simulation parameters are
listed in Table. \ref{table1}.

\begin{figure}[htbp]
    \centering
  
    \subfloat[Comparison of total energy con-\\sumption.]
    {\includegraphics[width=0.242\textwidth]{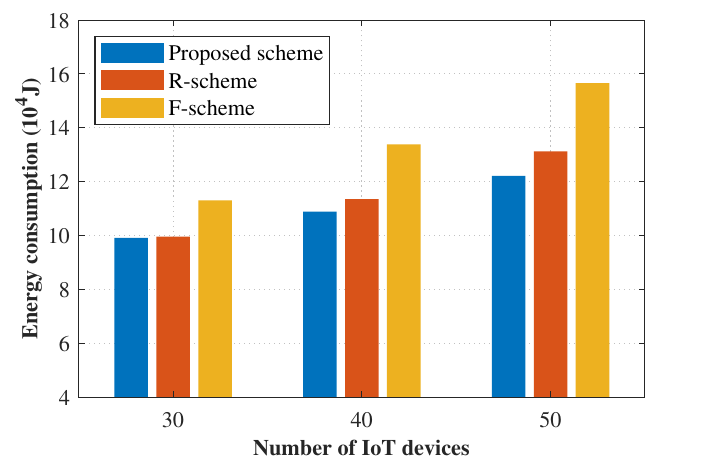}\label{fig:ee}}
    \hfill 
    \subfloat[Comparison of hovering energy consumption.]
    {\includegraphics[width=0.242\textwidth]{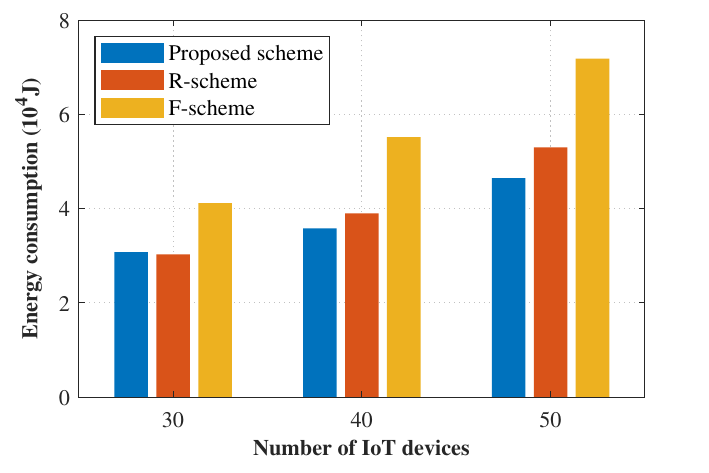}\label{fig:ff}}
    \hfill
    \subfloat[Comparison of UAVs flight dist-\\ance.]
    {\includegraphics[width=0.242\textwidth]{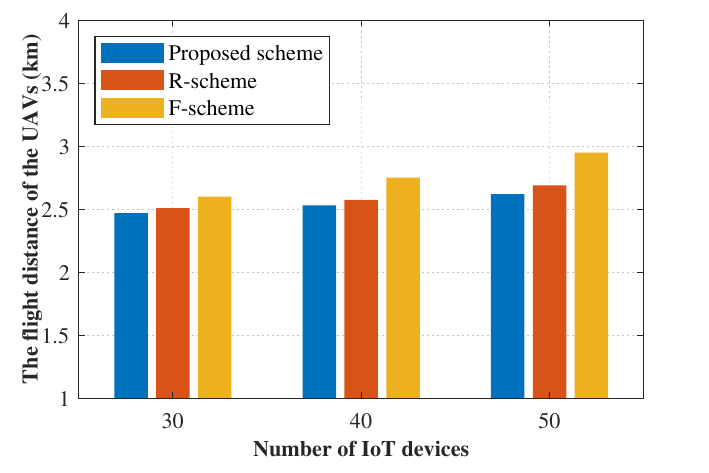}\label{fig:gg}}
    \hfill
    \subfloat[Comparison of UAV-LEO transmission time.]
    {\includegraphics[width=0.242\textwidth]{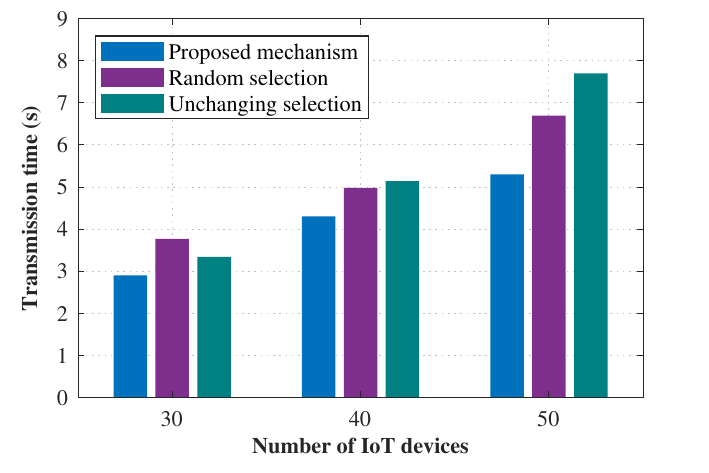}\label{fig:hh}}
    \hfill
    \caption{Comparison results of different algorithms.}
    \label{fig:nn}
  \end{figure}

\par To verify the superiority of our proposed algorithm for IoT-UAV communication, 
the simulation results are shown in Fig. \ref{fig:mm}\subref{fig:ee}-\subref{fig:gg}.
Two baseline approaches are introduced, marked as `R-scheme'  and `F-scheme' methods, respectively. 
In detail, in the `R-scheme' method, the IoT devices adopting NOMA technology are randomly paired without exchanging operations.
In the `F-scheme' method, all IoT devices evenly
share the resource block and the UAV hovers directly above the devices to collect data
without optimizing the hovering positions. The comparison results of energy consumption 
with different device counts are shown in Fig. \ref{fig:mm}\subref{fig:ee} and Fig. \ref{fig:mm}\subref{fig:ff}.
With the number of IoT devices growing, the energy consumption increases. Additionally, note that less energy is consumed compared with other 
benchmark algorithms, especially in the large-scale scenarios with more IoT devices.
This is because the growing number of IoT devices enables better performance of the exchange optimization operation in the proposed algorithm. 
Fig. \ref{fig:mm}\subref{fig:gg} illustrates the comparison results of UAVs flight distance with different number of IoT devices.
Clearly, the proposed scheme adopting the NOMA technology appears a shorter flight distance compared with other results. 
It is on account of the fact that the UAVs need to visit all the device 
nodes in the `F-scheme', resulting in the longer flight distance of 
UAVs and higher energy comsumption.
\par The comparison results of UAV-LEO
transmission delay with different numbers of IoT devices are analyzed in Fig. \ref{fig:mm}\subref{fig:hh}.  
Two benchmark mechanisms are introduced to evaluate the proposed LEO satellite selection mechanism. 
 In detail, in the `Random selection' mechanism, the UAV randomly selects a LEO satellite for offloading, 
 and in `Unchanging selection' mechanism, the UAV chooses an LEO  satellite for offloading until the satellite is no longer visible.
 It is observed that the proposed LEO selection mechanism can explore the best LEO satellite in real time,
and reduce the transmission delay compared with the other benchmark mechanisms.
Due to the time-varying LEO satellites, it is unrealistic to maintain the good channel conditions for 
communication links between UAVs and LEO satellites across the overall time period. Consequently, 
the transmission delay in the baseline schemes is excessively long, 
leading to more energy cost. In contrast, the proposed mechanism takes the 
high dynamics of LEO satellites into account, and UAVs can flexibly select the LEO satellite with the 
best channel condition, which significantly improves the performance.
\section{Conclusion\label{sec:conclusion}}
In this paper, we address the data collection and 
offloading problem by integrating UAVs trajectories planning and LEO satellite selection in SAGIN. 
The problem is formulated to minimize 
the energy consumption with the constraints on 
multi-dimension variables. Specifically, 
in the data collection phase from IoT to UAV, the algorithm
 is designed to optimize the IoT pairing, 
power optimization, UAV trajectory planning.
In the data offloading phase from UAV to LEO,
a real-time LEO satellite selection mechanism joint with STK is proposed.
Finally, simulation results verified
the effectiveness of the proposed approach, with about 10$\%$ less
energy consumption compared with the benchmark algorithm.

\textcolor{black}{\bibliographystyle{IEEEtran}
\bibliography{reference1.bib}
}

\end{document}